\documentclass[letter]{aa} 
\usepackage{color}

\DeclareRobustCommand{\rchi}{{\mathpalette\irchi\relax}}
\newcommand{\irchi}[2]{\raisebox{\depth}{$#1\chi$}} 

\newcommand{\rlight}{r_{\rm L}}

\usepackage{graphicx}
\usepackage{txfonts}
\usepackage[switch]{lineno}

\begin{document} 

\title{Localisation of the non-thermal X-ray emission of PSR~J2229+6114 from its multi-wavelength pulse profiles}

\author{
	J.~P\'etri\inst{1}
	\and 
	S.~Guillot\inst{2}
	\and
	L.~Guillemot\inst{3,4}
	\and
	D.~Mitra\inst{5,6}
	\and
	M.~Kerr\inst{7}
	\and
        L.~Kuiper\inst{8}
        \and
	I.~Cognard\inst{3,4}
	\and
	G.~Theureau\inst{3,4,9}
}
\institute{
    Universit\'e de Strasbourg, CNRS, Observatoire astronomique de Strasbourg, UMR 7550, F-67000 Strasbourg, France.\\
    \email{jerome.petri@astro.unistra.fr}  
    \and
    IRAP, CNRS, 9 avenue du Colonel Roche, BP 44346, F-31028 Toulouse Cedex 4, France      
    \and 
    Laboratoire de Physique et Chimie de l'Environnement et de l'Espace, Universit\'e d'Orl\'eans / CNRS, 45071 Orl\'eans Cedex 02, France
    \and
    Observatoire Radioastronomique de Nan\c{c}ay, Observatoire de Paris, Universit\'e PSL, Université d'Orl\'eans, CNRS, 18330 Nan\c{c}ay, France
    \and
    National Centre for Radio Astrophysics, Tata Institute for Fundamental Research, Post Bag 3, Ganeshkhind, Pune 411007, India
    \and
    Janusz Gil Institute of Astronomy, University of Zielona G\'ora, ul. Szafrana 2, 65-516 Zielona G\'ora, Poland 
    \and
    Space Science Division, Naval Research Laboratory, Washington, DC 20375-5352, USA
    \and
    SRON-Netherlands Institute for Space Research, Niels Bohrweg 4, 2333 CA, Leiden, The Netherlands
    \and
    LUTH, Observatoire de Paris, Universit\'e PSL, Universit\'e Paris Cit\'e, CNRS, 92195 Meudon, France
}

\date{Received ; accepted }

  \abstract
   {Pulsars are detected over the whole electromagnetic spectrum, from radio wavelengths up to very high energies, in the GeV-TeV range. Whereas the radio emission site for young pulsars is well constrained to occur at altitudes about several percent of the light-cylinder radius and $\gamma$-ray emission is believed to be produced in the striped wind, outside the light-cylinder, their non-thermal X-ray production site remains unknown.}
   {The aim of this letter is to localize the non-thermal X-ray emission region based on multi-wavelength pulse profile fitting for PSR~J2229+6114, a particularly good candidate due to its high X-ray brightness.}
   {Based on the geometry deduced from the joint radio and $\gamma$-ray pulse profiles, we fix the magnetic axis inclination angle and the line of sight inclination angle but we leave the region of X-ray emission unlocalised, somewhere between the surface and the light-cylinder. We localize this region and its extension by fitting the X-ray pulse profile as observed by the NICER, NuSTAR and RXTE telescopes in the ranges 2--7~keV, 3--10~keV and 9.4--22.4~keV, respectively.}
   {We constrain the non-thermal X-ray emission to arise from altitudes between $0.2\,\rlight$ and $0.55\,\rlight$ where $\rlight$ is the light cylinder radius. The magnetic obliquity is approximately $\alpha \approx 45\degr-50\degr$ and the line of sight inclination angle $\zeta \approx 32\degr-48\degr$.}
   {This letter is among the first works to tightly constrain the location of the non-thermal X-ray emission from pulsars. We plan to apply this procedure to several other good candidates to confirm this new result.}


\maketitle

%

\section{Introduction}
Among the almost 3,000~pulsars detected so far, a substantial fraction of the $\gamma$-ray emitters are radio-loud. For the subclass of young pulsars, the radio phenomenology is well constrained and understood as coming from an emission cone located at a height about several hundred kilometres above the stellar surface in the region where the dipole magnetic field dominates \citep{mitra_nature_2017}. These results are obtained from the radio polarization data interpreted using the rotating vector model \citep{radhakrishnan_magnetic_1969}. The high brightness temperatures of the radio emission require a collective or coherent emission mechanism. Based on the emission region constraint, coherent curvature radiation (CCR) by charge bunches is invoked as the plausible emission mechanism. Moreover, several works using either analytical methods, force-free electrodynamics (FFE) simulations or particle-in-cell simulations converged to the same $\gamma$-ray emission site, namely the current sheet of the striped wind outside the light-cylinder e.g. \citep{petri_unified_2011, petri_young_2021, cerutti_modelling_2016}. To produce these high energy photons incoherent curvature and synchrotron radiation are the two competing emission mechanisms. However, the non-thermal X-ray emission is less well understood and studied. There is currently no concrete evidence about its origin.

This letter aims at constraining the location of the non-thermal X-ray emission based on the geometry deduced from the combined radio and $\gamma$-ray pulse profile fitting. To apply this technique, good pulsar candidates should be seen in all three wavelength domains with a well defined pulse profile in the non-thermal X-ray range, above 1--2~keV, to avoid contamination from the surface thermal X-ray emission. We found that PSR~J2229+6114 satisfies all the requirements for our purpose. This pulsar was discovered by \citet{halpern_psr_2001} in radio and X-rays, pulsating at a period of $P \approx 51.6$~ms and with a spin-down rate of $\dot{P} \approx 7.83\times10^{-14}$~s/s. Its light-cylinder radius is about $\rlight = c/P = 2,462$~km where $c$ is the speed of light. 
Its typical surface magnetic field is therefore estimated from the magneto-dipole losses \citep{petri_theory_2016} to be around $2\times10^8$~T and its strength at the light-cylinder to $23$~T assuming a dipole-like radial decrease, $r^{-3}$. This pulsar later was detected by the \textit{Fermi} Large Area Telescope (LAT) instrument in the $\gamma$-ray energy range of [100~MeV,100~GeV] by \citet{abdo_fermi_2009-2}. They reported a phase lag between the strongest XMM-Newton X-ray peak in the 1--10~keV range and the peak in radio of about $0.17\pm0.02$. \cite{petri_unified_2011} already fitted the $\gamma$-ray pulse profile of PSR~J2229+6114 with a split monopole solution and more recently \citet{petri_young_2021} used a full 3D force-free magnetosphere model with a dipole field close to the surface. They found a magnetic obliquity of $\alpha \approx 35\degr$ and a line of sight inclination angle of $\zeta \approx 44 \degr$. In this Letter, we use joint radio and $\gamma$-ray data as well as new simulations with an emission height for the radio beam fixed at $R/\rlight=0.1$ as deduced from radio polarization observations and fitting with the rotating vector model including the aberration/retardation effects \citep{johnston_thousand-pulsar-array_2023}.

This Letter starts, in Section~\ref{sec:Data}, with some details about the data sets used in this work.  Then radio and $\gamma$-ray fitting results are summarized in Section~\ref{sec:Pulse} and extended by simultaneously joining the non-thermal X-ray pulse profile. The underlying expected particle dynamics and its energetics are discussed in Section~\ref{sec:Discussion}. Conclusions are drawn in Section~\ref{sec:Conclusion}.

\section{Multi-wavelength data sets}
\label{sec:Data}

\subsection{Radio and $\gamma$-ray pulse profiles}
\label{sec:radio_gamma}

\begin{figure}[h]
	\centering
	\includegraphics[width=\columnwidth]{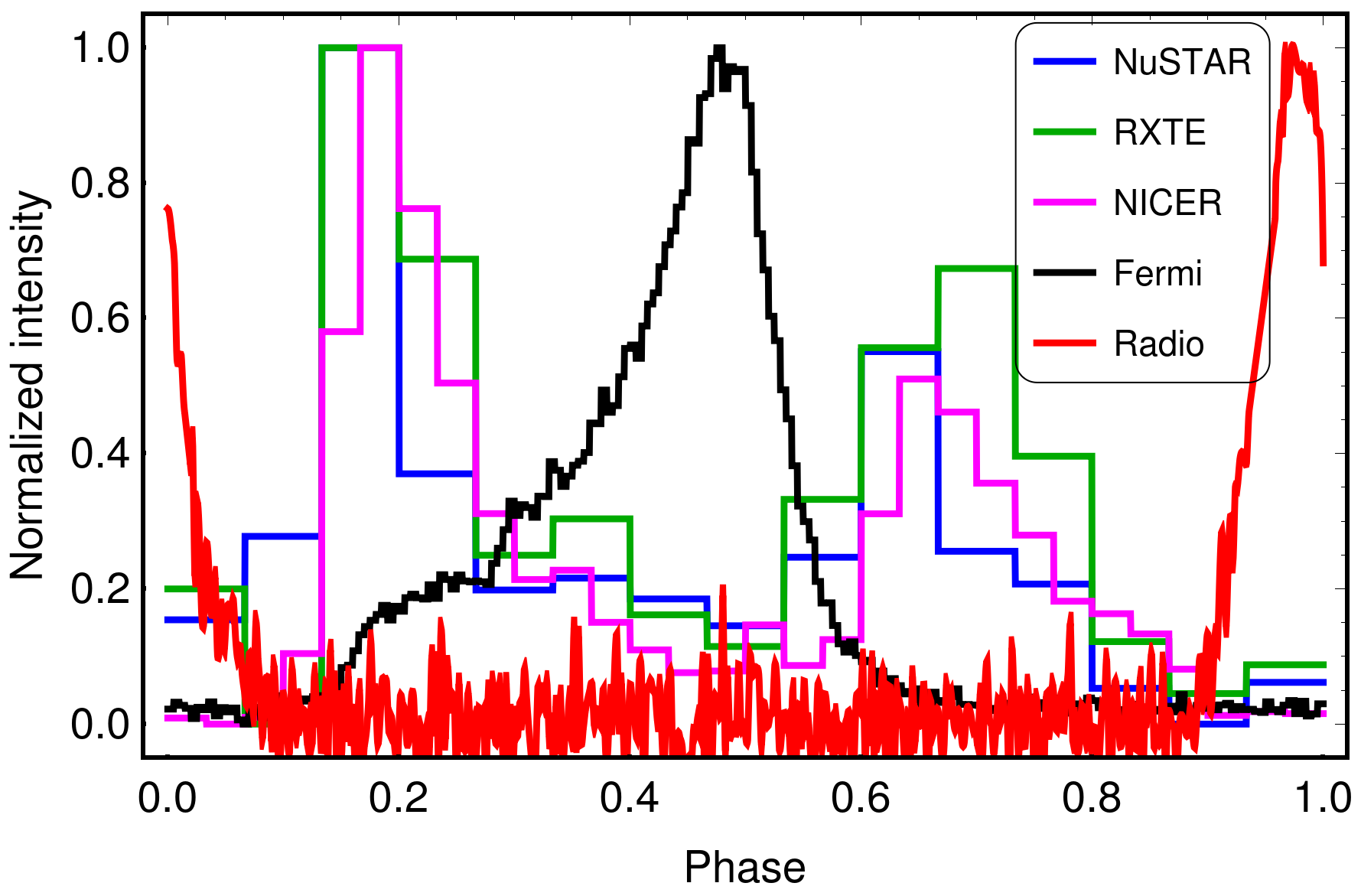}
	\caption{Multi-wavelength light curves of PSR~J2229+6114 as observed in radio with the Lovell telescope at the Jodrell Bank Observatory (1.5~GHz, red line), in X-ray with NuSTAR (3-10~keV, blue line), RXTE (9.4-22.4~keV, green line), NICER (2-6.9~keV, magenta line) and in $\gamma$ rays with the \textit{Fermi} LAT ($\geq100$~MeV, black line).}
	\label{fig:multilambdaj22296114}
\end{figure}

We start by investigating the radio and $\gamma$-ray pulse profiles using the  Fermi-LAT instrument, a wide field-of-view $\gamma$-ray telescope operating in the energy band 20~MeV to 300~GeV. \citep{atwood_evaluating_2099}.
Fig.~\ref{fig:multilambdaj22296114} summarizes the \textit{Fermi}-LAT pulse profile for energies above 0.1 GeV (black histograms), obtained by analyzing LAT data recorded between August 2008 and January 2023. The gamma-ray photon arrival times were phase-folded using a timing solution for PSR~J2229+6114 extracted from the same \textit{Fermi} LAT dataset, using the techniques presented in \cite{ajello_gamma-ray_2022}. The \textit{Fermi} LAT Third Catalog of Gamma-ray Pulsars \citep[3PC;][]{smith_third_2023} reported a double peaked $\gamma$-ray pulse profile showing a peak separation of $\Delta = 0.222\pm0.001$ and a $\gamma$-ray time lag of $\delta = 0.296\pm0.001$ between $\gamma$ and radio. The timing solution constructed for this work uses the same radio fiducial phase \citep[see e.g.,][for details]{smith_third_2023} as the timing solutions for this pulsar from the \textit{Fermi} LAT Second Catalog of Gamma-ray Pulsars \citep[2PC;][]{abdo_second_2013} and from 3PC, so that the radio/gamma-ray phase separation in Figure~\ref{fig:multilambdaj22296114} is consistent with that in the latter catalogs.  The 1.4~GHz radio profile in this Figure was taken from the 2PC auxiliary files archive, and corresponds to a profile recorded with the Lovell telescope at the Jodrell Bank Observatory (UK). 

\subsection{X-ray band}

We used several instruments in the X-ray band: NICER (Neutron star Interior Composition ExploreR, \citealt{gendreau_neutron_2012}), RXTE (Rossi X-ray Timing Explorer, \citealt{rothschild_-flight_1998}) and NuSTAR (Nuclear Spectroscopic Telescope Array, \citealt{harrison_nuclear_2013}),  together providing a good coverage from the soft to the hard X-rays (1--60~keV). The pulse profiles have also been phase-aligned with the radio profile, a crucial prerequisite for our model.

\subsubsection{NICER pulse profile}

The NICER observations of PSR~J2229+6114 from 2018-02-02 22:59:00 to 2022-12-29 06:55:40 (ObsID 1033370101--5033370272) were used to generate the X-ray pulse profile analysed here. The data were re-processed with the standard recommended procedure, using \texttt{nicerl2} of \texttt{NICERDAS} v10 and calibration file v20221001, distributed with \texttt{HEASOFT v6.31}. Following this, we employed the \texttt{NICERsoft} package\footnote{\url{https://github.com/paulray/NICERsoft}} to further filter the data, excluding periods with 1) Earth magnetic cut-off rigidity \texttt{COR\_SAX$<$1.5} GeV/c, 2) space weather index KP$\ge 5$, 3) overshoot rates larger than 1.5 c/s and 4) undershoot rates larger than 600 c/s.  We also remove observation periods with a 2--10~keV count rate $> 1$c/s (averaged over 32 s) to exclude possibly remaining periods of high background.  

After this careful filtering, the event files of the remaining ObsIDs were merged, and the phases of all detected events were calculated using \texttt{photonphase} (from \texttt{PINT}\footnote{\url{https://github.com/nanograv/PINT}}). The phase-folding was carried out using the timing solution for PSR~J2229+6114 extracted from \textit{Fermi} LAT data, as mentioned in Section~\ref{sec:radio_gamma}, whose validity interval encompasses the NICER dataset. The NICER pulse profile is presented in Figure~\ref{fig:multilambdaj22296114} (purple histogram) and \ref{fig:j22296114} (top panel), and obtained in the 1.05--6.9 keV energy interval (after optimization, following \citealt{guillot_nicer_2019}).


\subsubsection{RXTE-PCA pulse profile}
The hard X-ray profile (9.4--22.4~ keV) used in this work (see Figure~\ref{fig:multilambdaj22296114} (green histogram) and \ref{fig:j22296114} (bottom panel)) is adopted from \citet{kuiper_soft_2015} who showed RXTE-PCA profiles of PSR~J2229+6114 for two different energy bands, 1.9--9.4~keV and 9.4--22.4~keV, from a combination of two observation runs totalling an exposure time of 220.2 ks. More detailed information on the RXTE PCA analysis can be found in Section~5.18 of \citet{kuiper_soft_2015}.

\subsubsection{NuSTAR pulse profile}
The hard X-ray telescope NuSTAR \citep[][]{harrison_nuclear_2013}, operating in the 3--79~keV range, observed PSR~J2229+6114 and its pulsar wind nebula 'the Boomerang' for about 45.2~ks on Sept. 21--22, 2020 (MJD 59114.012-59114.99).
We used the `default' screened event files and converted the event arrival times to Solar System Barycentre arrival times, adopting Solar System ephemeris DE405, and NuSTAR clock correction file {\it nuCclock20100101v137.fits} yielding time tags with an absolute timing accuracy of better than $100 \mu$s \citep[][]{bachetti_timing_2021}, amply sufficient to study the timing signal of PSR~J2229+6114. Next, events were selected from a circular extraction radius of $90\arcsec$ centered on the pulsar. The radio-aligned pulse profile for events with measured energies in the 10--60~keV band (= RXTE PCA analogon) is shown in Fig.~\ref{fig:multilambdaj22296114} (blue histogram). The $Z_n^2$ significance, specifying the deviation from uniformity, is about $8.3 \sigma$ adopting 8 harmonics ($n=8$).
For the spectral energy distribution in X-ray and $\gamma$-ray we refer for instance to Fig.~28 of \cite{kuiper_soft_2015}, Fig.~2 of \cite{iniguez-pascual_synchro-curvature_2022} and in fig.~3 of \cite{cotizelati_spectral_2020}.

\section{Multi-wavelength pulse profile fitting}
\label{sec:Pulse}
Investigating simultaneously the radio, X-ray and $\gamma$-ray pulse profiles allows constraining the most probable value of the magnetic obliquity and line of sight inclination angle. Localizing the height of the X-ray emission site is another goal of this study. In this section we show a self-consistent view of the pulsar geometry in agreement with the three pulse profiles. We proceed in two steps. In a first stage, we fit the $\gamma$-ray profile and its time lag with respect to the radio peak to extract the magnetic obliquity~$\alpha$ and line of sight inclination angle~$\zeta$. In a second stage we add the X-ray emission zone, allowing emission to start above the radio emission site at $0.2\,\rlight$ to almost the light-cylinder in order to deduce the extension and altitude of this new region. 

\subsection{Radio and $\gamma$-ray emission}

\begin{figure}[h]
	\centering
	\includegraphics[width=\columnwidth]{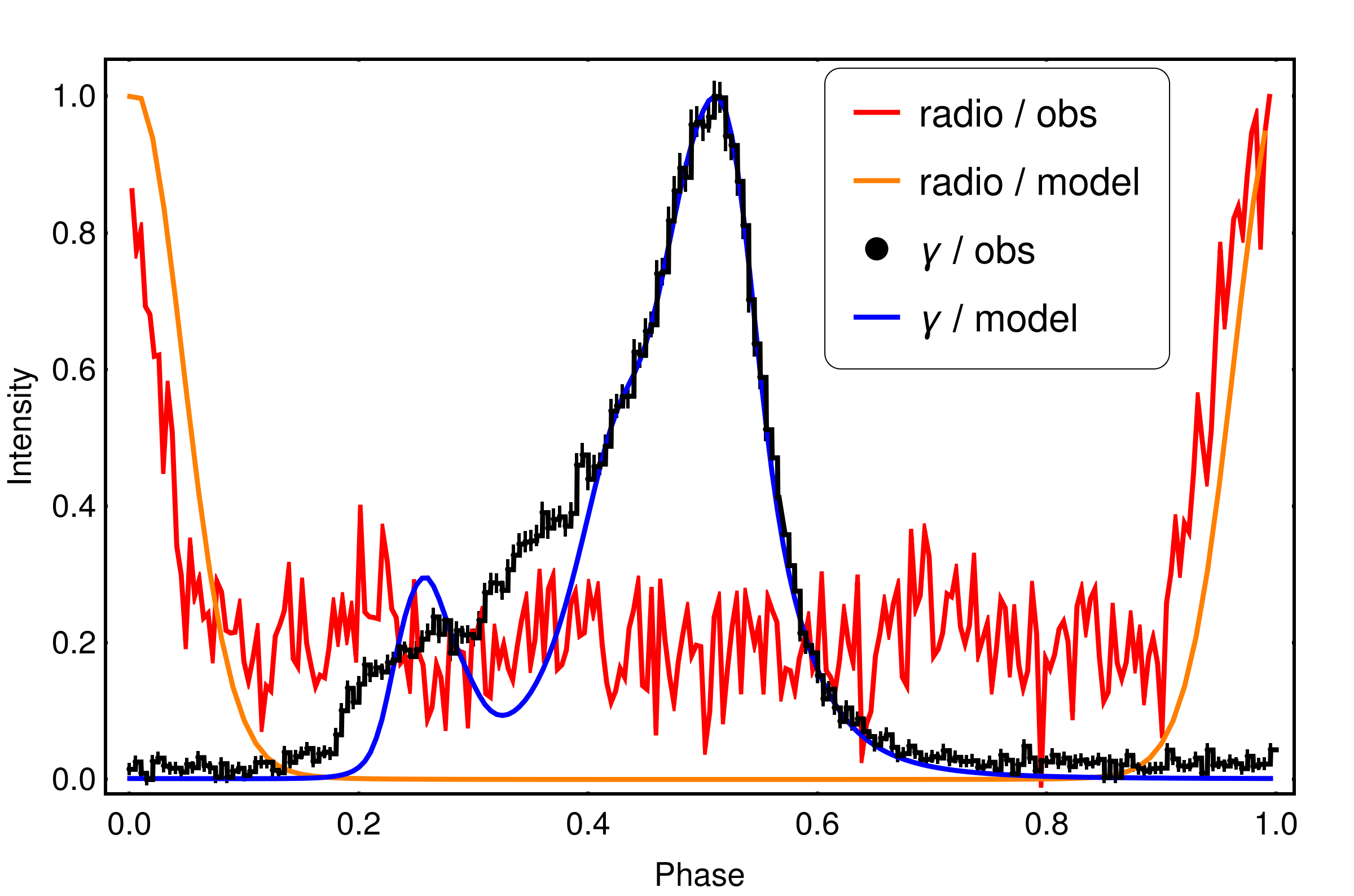}
	\caption{Best fit for the phase-aligned $\gamma$-ray light-curve ($\geq100$~MeV) of PSR~J2229+6114 with $(\alpha,\zeta)=(45\degr, 38\degr)$. The radio pulse profile is shown in red, the model in orange, the $\gamma$-ray observations in black and our model in blue.}
	\label{fig:fitj222961141l0}
\end{figure}

PSR~J2229+6114 was studied more than a decade ago by \citet{petri_unified_2011}, who found $\alpha=45\degr$ and $\zeta=40\degr$ assuming a simple split monopole current sheet geometry. More recently, \citet{petri_young_2021} found $\alpha=35\degr$ and $\zeta=44\degr$ with an accurate dipole force-free magnetosphere numerical solution. 
Here we reanalyse the latter model, from the latest force-free magnetosphere simulations \citep{petri_multi-wavelength_2024} with $r/\rlight=0.1$. The accuracy of the obliquity~$\alpha$ will be within $\Delta\alpha=5\degr$ because simulations have been performed with this angle increment $\Delta\alpha$. Because we detect both radio and $\gamma$-ray photons, the condition $\alpha \approx \zeta$ must apply within an error bar related to the opening angle of the radio emission cone~$\delta_{\rm rc}$. From \citet{petri_unified_2011} we know that $\cos(\pi\,\Delta) = |\cot \alpha \, \cot \zeta|$, and thus $\cos(\pi\,\Delta) = \cot^2 \alpha$. Therefore the obliquity is around $\alpha \approx \zeta \approx 48\degr$.
Figure~\ref{fig:fitj222961141l0} shows the best fit by minimizing the residual defined by
\begin{equation}
	\label{eq:chi_square}
    \rchi^2_{\nu} = \frac{1}{n-1} \sum_{i=1}^n {\left( I_i^{\rm mod} - I_i^{\rm obs} \right)^2}/{ \sigma_i^2} 
\end{equation}
where $n$ is the number of data points, $\nu=n-1$, $I_i^{\rm obs}$ the observed $\gamma$-ray flux and $I_i^{\rm mod}$ the predicted $\gamma$-ray normalised intensity, and $\sigma_i$ the uncertainties in the gamma-ray flux.
As best fit, we chose an inclination of the magnetic dipole of $\alpha=45\degr$ and an observer line of sight angle of $\zeta=38\degr$. These angles are deduced by minimizing~$\rchi^2_{n-1}$ in eq.~\eqref{eq:chi_square}. The fit is excellent for the strongest peak but less good for the weak peak that is more pronounced in the energy band 50~MeV-300~MeV. Moreover knowing the radio pulse profile width~$W$ we can estimate the radio emission cone half-opening angle $\delta_{\rm rc}$ from \citet{gil_geometry_1984}:
\begin{equation}
	\sin (\delta_{\rm rc}/2) = \sqrt{ \sin^2 (W/4) \, \sin \alpha \, \sin ( \alpha + \beta ) + \sin^2 (\beta/2) } .
\end{equation}	
The observed width~$W$ assumes that the full cone is visible at each rotation.
The line of sight must then satisfy $|\alpha-\zeta|<\delta_{\rm rc}$ in order to cross the radio beam. With the above fit we get for PSR~J2229+6114 an opening angle of $\delta_{\rm rc} \approx25\degr$. Assuming a dipole field structure, the radio emission height becomes $h/\rlight \approx (4 / 9) \, \delta_{\rm rc}^2 \approx 0.084$ therefore slightly less than 10\% of the light-cylinder, consistent with the emission height estimated from radio polarization measurements \citep{mitra_nature_2017, johnston_thousand-pulsar-array_2023}.

\subsection{X-ray emission}

\begin{figure}[h]
	\centering
	\includegraphics[width=\columnwidth]{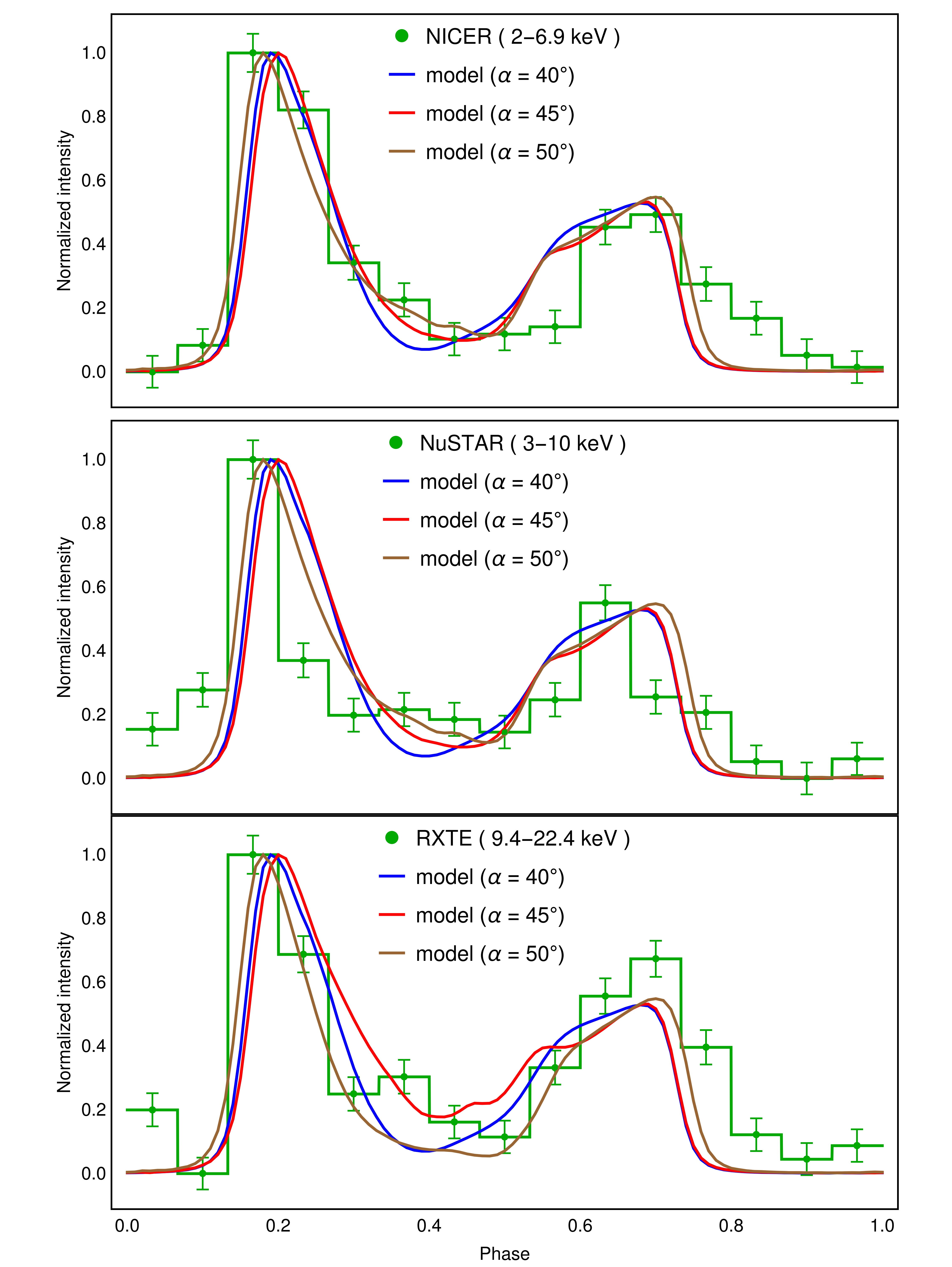} 
	\caption{Best fitted light-curves in X-rays using the NICER data, in the upper panel, NuSTAR data in the middle panel and the RXTE data, in the lower panel. Observations are shown in green and the models in blue for $(\alpha,\zeta) = (40\degr, 46\degr)$, in red for $(\alpha,\zeta) = (45\degr, 38\degr)$ and in brown for $(\alpha,\zeta) = (50\degr, 36\degr)$.}
	\label{fig:j22296114}
\end{figure}

In order to deduce the X-ray emission altitude and extension, we computed several atlases of pulse profiles assuming that photons emanate from thin shells around the separatrix, i.e. the surface representing the interface between closed and open magnetic field lines. We considered 15~shells with boundaries along magnetic field lines, an emissivity with a Gaussian profile in the direction perpendicular to the magnetic surface of width $w_X=0.1\rlight$, and sharp cut offs at radii $r_1$ and $r_2$ such that a shell is in the interval $[r_1,r_2]$. These zones possess an extension in radius of $0.05\,\rlight$, cut in different spherical layers from a first zone with height in $r/\rlight\in[0.2,0.25]$ to a last zone at height $r/\rlight\in[0.9,0.95]$ therefore the $k$-th zone is $(r/\rlight)_k \in [0.2+0.05\,k,0.25+0.05\,k]$, starting from zone $k=0$ to $k=14$. From these atlases, we compute pulse profiles by using only one zone or by adding several adjacent zones to determine the minimal and maximal emission heights compatible with the pulses. We also kept the line of sight inclination angle~$\zeta$ as a free parameter to check whether we retrieve the same value as for the $\gamma$-ray light curve. For completeness, an obliquity with $\alpha=50\degr$ was also tested. Performing the fits for NICER above 2~keV, for NuSTAR above 3~keV, and for RXTE in the band $[9.4,22.4]$~keV, we get the results shown in Figure~\ref{fig:j22296114}. The top panel compares the NICER pulses in green to our model in blue for $\alpha=45\degr$ and in red for $\alpha=50\degr$, giving a best fit of $\zeta=46\degr$ for the blue curve and $\zeta=32\degr$ for the red curve. The middle panel shows good fits for NuSTAR with $\zeta=48\degr$. The bottom panel shows the best fit for RXTE in green compared to our model in red and blue, giving a best fit of $\zeta=48\degr$ and $\zeta=34\degr$ respectively. These values of $\zeta$ encompass the value found in the previous paragraph and are fully compatible. 
In all cases, the emission starts at $r/\rlight=0.2$ and goes up to an altitude of $r/\rlight=0.55$ for all three panels in Figure~\ref{fig:j22296114}. The results are summarized in Table~\ref{tab:Xray_geometry} with the associated residual~$R$ in the better-fitting geometry range $R\in[1.15,1.37]$. We used a consistent set of obliquities~$\alpha$ and line of sight inclination~$\zeta$ for all light curves shown in Figure~\ref{fig:j22296114} and corresponding to the best fit for NICER observations.

We stress that our force free model only provides the geometrical locations where the radiation is produced within the pulsar magnetosphere. The force-free simulations used in this work assume a star-centred dipole, and hence if the non-thermal X-rays originate from below the radio emission region the model predicts time aligned radio and non-thermal X-ray profiles, which is not consistent with the observations. 
Clearly the expected width below the radio emission region is significantly smaller than the observed X-ray profile width. See also \cite{philippov_origin_2020}, who suggest a different explanation about the particles producing the radio emission. Realistically the surface magnetic field is non-dipolar in nature \citep{petri_joint_2020, arumugasamy_evaluating_2019}, and while this can explain the observed radio and non-thermal X-ray profile misalignment, this cannot explain the observed X-ray profile width since the presence of a non-dipolar field will shrink the polar cap further resulting in a smaller pulse width compared to the dipolar case. The fact that the non-thermal X-ray emission originates higher up in the magnetosphere is therefore a robust result.
\begin{table}[h]
	\centering
	\begin{tabular}{ccccc}
		\hline
    Satellite & Band & $\alpha=40\degr$ & $\alpha=45\degr$ & $\alpha=50\degr$ \\
		\hline
		NICER & 2--6.9~keV & $46\degr$ & $38\degr$ & $36\degr$ \\
		RXTE & 9.4--22.4~keV  & $58\degr$ & $48\degr$ & $34\degr$ \\
		NuSTAR & 3--10~keV  & $58\degr$ & $56\degr$ & $50\degr$ \\
		\hline
	\end{tabular}
\caption{Line of sight inclination angles~$\zeta$ 
	implied by the non-thermal X-ray pulse profile obtained with NICER, RXTE and NuSTAR data by imposing $\alpha = \{ 40\degr, 45\degr, 50\degr \}$.
	\label{tab:Xray_geometry}}
\end{table}

\section{Discussion on the energetics}
\label{sec:Discussion}
In our force free model, the region along the separatrix is the location where the non-thermal X-rays are generated. The charges that accelerate along the separatrix give rise to non-thermal X-ray emission by two possible mechanisms, namely, synchrotron or curvature radiation. The former requires a large pitch angle~$\psi \sim 90\degr$ for the particles to significantly experience the magnetic gyro-motion. The latter requires motion along magnetic field lines with a possible perpendicular drift. This means a small pitch angle $\psi\ll1$. For intermediate pitch angles, neither synchrotron nor curvature spectra applies but synchro-curvature sets in. 
This synchro-curvature  radiation has been studied in the context of pulsar magnetospheres by \citet{kelner_synchro-curvature_2015} where particles are forced to glide along the separatrix. 

Let us compute the expected particle characteristics in the force-free approximation. The magnetic field strength of the dipole is known from its spin-down and the field line curvature from the force-free model. Assuming a dipolar decrease of the magnetic field strength from the surface to the base of the wind, its strength at the light-cylinder is about $B_{\rm L} \approx 23~T$. In order to produce the required X-ray photons of energy $E_{\rm X}$, typically above 1~keV, the particle Lorentz factor must be the following, in the curvature~$\gamma_{\rm curv}$ and synchrotron $\gamma_{\rm sync}$ case, respectively:
\begin{subequations}
	\begin{align}
		\gamma_{\rm curv} & 
		\approx 3.5\times10^5 \, \left( {E_{\rm X}}/{5~\textrm{keV}} \right)^{1/3} \, \left( {\rho_{\rm c}}/{\rlight}\right)^{1/3} \\
		\gamma_{\rm sync} & 
		\approx 1358 \, \left( {E_{\rm X}}/{5~\textrm{keV}} \right)^{1/2} \, \left( {B}/{B_{\rm L}}\right)^{-1/2}
	\end{align}
\end{subequations}
where $\rho_{\rm c}$ is the curvature radius of the particle trajectory, $B_{\rm L}$ the magnetic field strength at the light-cylinder, $B$ the magnetic field strength at the emission location. However, synchrotron radiation is very unlikely close to the stellar surface because particles stay in their fundamental Landau level. At larger distances, close to the light cylinder, this pitch angle could be measured from the radiation reaction limit velocity as explained by \cite{kalapotharakos_fundamental_2019}.

\begin{figure}[h]
	\centering
	\includegraphics[width=0.9\columnwidth]{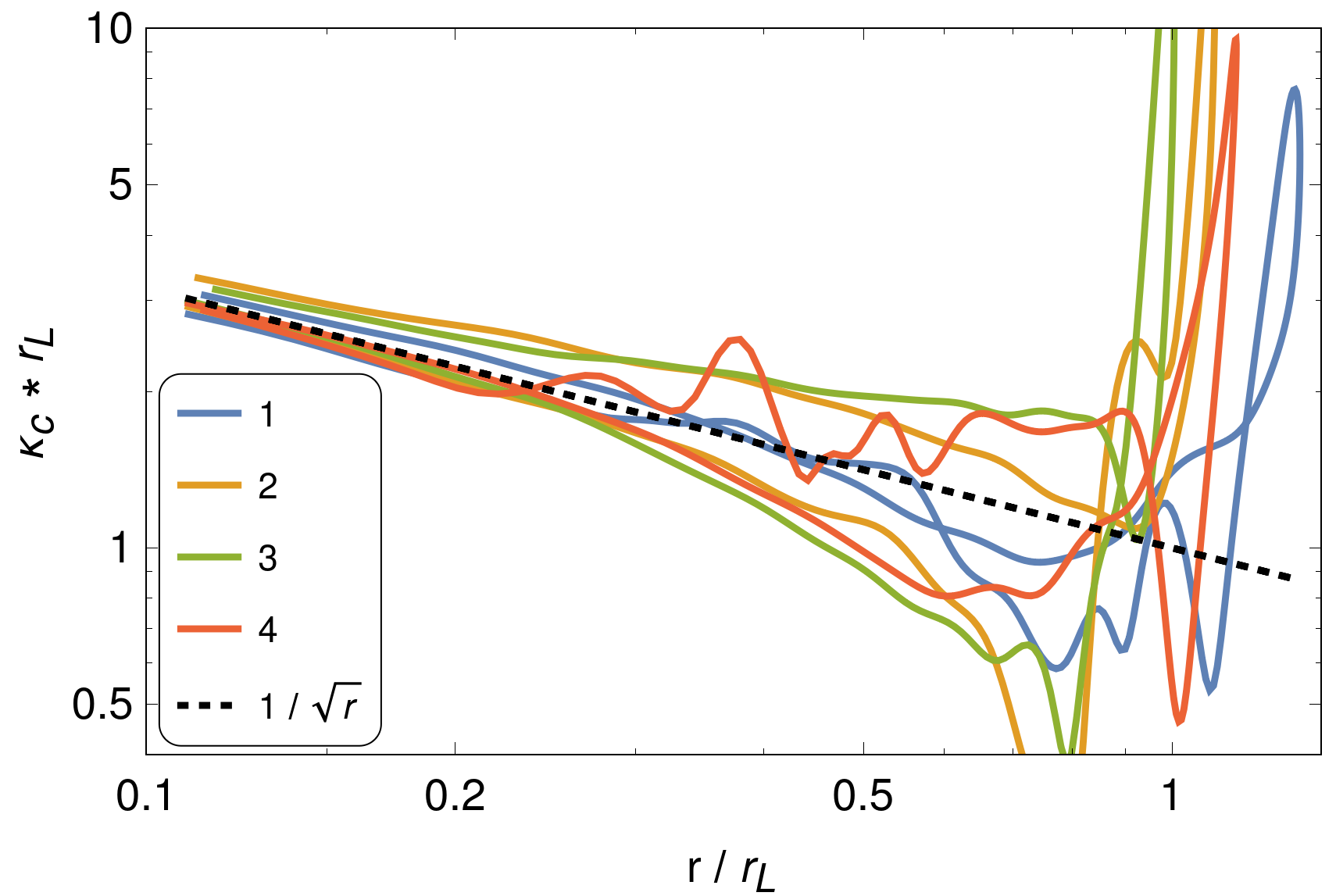}
	\caption{Curvature in units of $1/\rlight$ along the separatrix starting from the surface going to the light-cylinder and back to the surface for the force-free solution. Values are shown for an oblique rotator with $\alpha=45\degr$. At small heights it behaves as $\kappa_{\rm c} \, \rlight \approx r^{-1/2}$. Four field lines are shown in different colours.}
	\label{fig:courbure_separatrice_vide_ffe}
\end{figure}

Moreover, from a theoretical point of view, in the pair cascade region above the polar cap, current models predict two populations of pair plasma flows: a primary beam $(b)$ of electrons/positrons with very high Lorentz factor accelerated in the vacuum gap potential drop to reach $\gamma_{\rm b} \approx 10^6$ and a secondary plasma of $e^\pm$ pairs $(p)$ produced by cascades due to magnetic photo-disintegration reaching Lorentz factors of $\gamma_{\rm p} \approx 10^2$ \citep{kazbegi_circular_1991, arendt_pair_2002, usov_two-stream_2002}. In the partially screened gap model of \citet{gil_drifting_2003}, an ion outflow is allowed with $\gamma_{\rm ion}\approx 10^3$. Electrons and positrons do not follow exactly the same distribution functions because of the parallel electric field screening \citep{beskin_physics_1993}. The pair multiplicity factor can reach values up to $\kappa \approx 10^4-10^5$ \citep{timokhin_maximum_2019}. This pair plasma produces the observed radio emission, typically in the MHz--GHz band, through CCR \citep{mitra_nature_2017}. The evolution of the curvature~$\kappa_{\rm c}$ along the particle path within the separatrix surface in a force-free magnetosphere plays a central role. It is shown in Fig.~\ref{fig:courbure_separatrice_vide_ffe} for an obliquity of $\alpha=45\degr$, in units of $1/\rlight$. Each colour depicts a different field line, a sample of 4 representative lines having been chosen. The curvature decreases with distance to the star approximately like $r^{-1/2}$ for the FFE model (actually it is the same behaviour for the vacuum case), as long as $r\ll\rlight$.
Moreover at the radio emission height of $r/\rlight \approx 0.1$, this curvature is about $\kappa_{\rm c} \, \rlight = \rlight / \rho_{\rm c} \approx 22-35$. Therefore the particle Lorentz factor for producing radio photons in the GHz band is
$\gamma_{\rm radio} \approx 57 \, \left( \nu_{\rm radio} / 1~\textrm{GHz} \right)^{1/3} \, \left( 30 \, \rho_{\rm c} / \rlight \right)^{1/3}$ 
which corresponds to the secondary plasma flow with $\gamma_{\rm radio} \approx 100$ for typical values of $\rho_{\rm c}$. If the X-ray photons are produced in regions with similar curvature, then the Lorentz factor ratio must be
$\gamma_{\rm X}/ \gamma_{\rm radio} = \left( 5~\textrm{keV} / h\times1~\textrm{GHz} \right)^{1/3} \approx 10^3$ 
which corresponds to the primary beam with Lorentz factor of about $10^{4-5}$, consistent with the estimates for primary beams, see for instance \cite{timokhin_current_2013}. However the core of the X-ray emission approaches $r\approx 0.5 \, \rlight$ where the curvature has decreased by a factor two to three $\kappa_{\rm c} \, \rlight \approx 10$ requiring 
$\gamma_{\rm X} \approx 1.6 \times 10^5 \, \left( E_{\rm X}/ 5~\textrm{keV} \right)^{1/3} \, \left( 10 \, \rho_{\rm c}/ \rlight \right)^{1/3}$. 
Therefore radio and non-thermal X-ray emission are produced by CCR of the secondary plasma ($\gamma_{\rm b} \approx 100$) and the incoherent curvature radiation of the primary beam ($\gamma_{\rm p}\gtrsim10^5$) respectively. This is consistent with the one-dimensional particle distribution functions of the out-flowing relativistic plasma along the open magnetic field lines. 
Above a height of $0.5 \, \rlight$ the sharp decrease of the curvature~$\kappa_{\rm c}$ shifts the photon energy to a lower band well below 1~keV. Moreover, the particle density number also drops due to the divergent magnetic field structure and the spherical expansion.

Non dipolar components are excluded in the radio emission region and above it. The reason why non-thermal X-rays are mostly produced along the separatrix is two fold. First, the curvature radiation power scales as the curvature squared $\kappa_{\rm c}^2$  and the particle charge squared $q^2$ such that $\mathcal{P}_{\rm c} = (q^2/6\,\pi\,\varepsilon_0) \, \gamma^4 \, c \, \kappa_{\rm c}^2$. Because the curvature $\kappa_{\rm c}$ drastically decreases towards the centre of the polar cap, the associated curvature power also decreases, even faster than $\kappa_{\rm c}$. On top of that, polar caps are known to show pair creation only in regions where the current density $j$ satisfies $j>\rho\,c$ or $j/\rho\,c<0$, the physical boundaries depending on the obliquity \citep{timokhin_current_2013} and $\rho$ being the corotating charge density. Moreover, magnetic field lines near the poles do not sustain pair cascades, producing only low energy primary particles. This leads to a hollow cone model reminiscent of the radio hollow cone model. Second, the particle density number~$n_e$ along the separatrix is high due to the current required to support the transition layer between the open field line region and the closed field line region. Moreover, the current density decreases towards the centre for an inclined rotator and vanishes for an orthogonal rotator, see Figures~8 and 9 of \citet{petri_spheroidal_2022}. Because the emissivity is proportional to the product $n_e\,\mathcal{P}_{\rm c}$, we expect the light-curve to be formed essentially in the separatrix region as postulated in our model. Adding some small resistivity to create a parallel accelerating electric field would only slightly change the value of the curvature~$\kappa_{\rm c}$. All the above results would be essentially unchanged for a resistive magnetosphere.

\section{Conclusion}
\label{sec:Conclusion}
Based on the multi-wavelength light-curve fitting of PSR~J2229+6114, we showed that the non-thermal X-ray photons emanate from a region located between the polar cap and the light-cylinder, along the separatrix, at an altitude in the range $r/\rlight\in[0.2,0.55]$. Curvature radiation of the primary beam with Lorentz factor $\gamma_{\rm p} \gtrsim10^5$ is responsible for this non-thermal X-ray emission whereas the secondary plasma with $\gamma_{\rm b} \approx100$ radiates radio photons. The extension of the X-ray region is controlled by the decrease in the open field line curvature along the separatrix, shifting the photon energy well below the X-ray band. 

\begin{acknowledgements}
We are grateful to P. Arumugasamy for helpful discussions and suggestions. 
This work has been supported by the CEFIPRA grant IFC/F5904-B/2018 and ANR-20-CE31-0010. SG acknowledges the support of the CNES. DM acknowledges the support of the Department of Atomic Energy, Government of India, under project No. 12-R\&D-TFR-5.02-0700. The \textit{Fermi}-LAT Collaboration acknowledges support for LAT development, operation and data analysis from NASA and DOE (United States), CEA/Irfu and IN2P3/CNRS (France), ASI and INFN (Italy), MEXT, KEK, and JAXA (Japan), and the K.A.~Wallenberg Foundation, the Swedish Research Council and the National Space Board (Sweden). Science analysis support in the operations phase from INAF (Italy) and CNES (France) is also gratefully acknowledged. This work performed in part under DOE Contract DE-AC02-76SF00515. Work at NRL is supported by NASA.
\end{acknowledgements}


\end{document}